# Feshbach Resonances and Limiting Universal Thermodynamics of Strongly Correlated Nucleons

A.Z.Mekjian
Department of Physics and Astronomy, Rutgers University, Piscataway, NJ 08854

Abstract
*A finite temperature model of strongly correlated nucleons with underlying isospin symmetries is developed. The model can be used to study the role of bound states and Feshbach resonances on the thermal properties of a spin ½, isospin ½ system of protons and neutrons by varying the proton fraction. An analysis of features associated with a universal thermodynamic limit or unitary limit is given. In the limit of very large scattering length, the effective range to quantum thermal wavelength appears as a limiting scale in an interaction energy and equation of state.*

PACS 24.10.Pa, 21.65.Mn, 21.65.Cd, 67.85.Pq, 03.75.-b

Keywords: Two specie Fermi gases, unitary limit, Feshbach resonances, precursor liquid/gas phase transition

*Introduction.* The behavior of dilute Fermi systems is of recent interest in both atomic systems and in nuclear systems. Early theoretical interest arose when Bertsch [1] formulated a many body challenge problem for systems with large scattering length compared to interparticle spacings. This regime is referred to as the unitary limit. Experimental studies of cold gases in the unitary regime in atomic physics were carried out by several groups [2-5]. In atomic systems Feshbach resonances have provided a unique tool for studies of such systems. In particular, tuning across a Feshbach resonance via a magnetic field has been a very successful method in the study of the transition from a Bose-Einstein condensate of tightly bound dimers to a BSC superfluid state. Some early theoretical studies of atomic systems were carried out by Ho and collaborators both at $T = 0$ [6] and $T \neq 0$ [7]. Moreover, results in ref.[7] showed that several of the features of a degenerate $T = 0$ Fermi gas are present in the high temperature Boltzmann limit. In nuclear physics, a Monte-Carlo investigation of the superfluid properties of a system of pure neutrons was done in Ref.[8]. Further theoretical understanding of the pure neutron system came from the extensive work of Bulgac and collaborators [9] and references therein. Early work on the Bertsch problem was done by Barker [10] and Heiselberg [11]. The inner crust of a neutron star is an example where dilute Fermi gases occur [12].

The focus of this paper centers around thermodynamic properties of strongly correlated fermions and in particular to a two component hadronic system made of protons and neutrons each with two spin states and underlying isospin symmetries. Real hadronic systems contain both neutrons and protons even in the limit of neutron stars which contain a small fraction of protons. Moreover, future FRIB (Facility for Rare Isotope Beams) experiments will study properties of nuclei with large neutron and proton excess. An investigation of a two component system is a rich extension of results from a one component system. For example, in a system of protons and neutrons both isospin symmetries and features associated with spin structure come into play. The n-p system has a bound state in the spin $S = 1$, isospin $I = 0$ state, while the n-n $S = 0, I = 1$ system and n-p $S = 0, I = 1$ have resonant like virtual states with very large scattering lengths. Also, the formation of deuterons in a dilute Fermi system [13] is a precursor to the liquid-gas phase transition [14]. In a gas to liquid phase transition clusters of all sizes appear. The specific heat shows a singular behavior around the first order liquid/gas phase transition [15]. While the nuclear system does not allow tuning with a magnetic field, tuning can be done by varying the proton fraction. Temperature also plays an important part in the number of bound and virtual states.

*Thermodynamic properties of the interacting gas with two components.* The equation of state EOS around a non-degenerate limit has a virial expansion that is $PV/k_B T = A - (x_2/x_1^2)A^2 + ((4x_2^2 - 2x_1 x_2)/x_1^2)A^3 + ....$ with $x_k$ correlation or cluster coefficients. For a one component system of identical fermions of spin $S = ½$, antisymmetrization effects result in $x_k = (-1)^{k+1}/k^{5/2}(V/\lambda_T^3)$. The thermal wavelength is $\lambda_T = h/\sqrt{2\pi m k_B T}$. A virial expansion is valid when $(A/V)\lambda_T^3$ is small. The above results can be extended to include interactions by considering the following model with two components made of protons and neutrons. The pressure in a dilute interacting gas to second order in the density is $PV/k_B T = A - b_2 A^2$ where the coefficient $b_2 \sim 1/V$ has contributions from $nn, pp$ and $np$ components. To see the structure of $b_2$ a simple example will be given where all three systems can form an $s-$wave bound state, with the $np$ system having two possibilities with spin $S = 1$, the deuteron, or $S = 0$. This will then be corrected for continuum interactions and the $nn$ and $pp$ bound states will be turned off. The law of partial pressures leads to a total pressure $PV/k_B T = N_p + N_n + N_{nn} + N_{pp} + N_{np}$ The $Z = N_p + N_{np} + 2N_{pp}$ and $N = N_n + N_{np} + 2N_{nn}$ and $A = Z + N$. The $N_{ij}(S)$ is

$$N_{ij}(S) = \frac{2S+1}{2^2} 2^{3/2} \frac{\lambda_T^3}{V} \exp[\frac{E_B(ij,S)}{k_B T}] N_i N_j \equiv a_2(ij,S) N_i N_j \tag{1}$$

with $i = p, n$ & $j = p, n$. The above law expresses chemical equilibrium[13]. The $N_{ij}(S)$ is strongly T dependent with higher $T$ breaking the bound state into its constituents. The $S$ is the spin of the $(ij)$ pairs and $E_B(ij)$ equals the binding energy of the pair. Writing $N_p = Z - N_{np} - 2N_{pp}$, $N_n = N - N_{np} - 2N_{nn}$, proton fraction $y_p = y = Z/A$ and neutron fraction $y_n = N/A = 1 - y_p = 1 - y$ leads to an equation of state EOS to order $A^2$ that is

$$\frac{PV}{k_B T} = A + \frac{1}{2^{5/2}} \frac{\lambda_T^3}{(2S+1)V} (y^2 + (1-y)^2) A^2 - \sum_S \sum_{i \& j = p, n} y_i y_j a_2(ij, S) A^2 \tag{2}$$

Anti-symmetrization corrections in the $pp$ and $nn$ channels are included in this equation. As noted the $pp$, $nn$ and $np$ $S = 0$, $I = 1$ channels have no bound states. However a resonance like virtual state acts as a bound state and makes a contribution to the $A^2$ term through a term due to Beth and Ulhenbeck[16-18] that changes the bound state

Boltzmann factor to a continuum correlation integral. Specifically,
$Exp[E_B/k_BT] \to (1/\pi)\int(\partial\delta_0/\partial k)\exp(-\hbar^2k^2/2\mu k_BT)dk$ where $\delta_0$ is the $s-$wave phase shift and $\mu = m/2$ is the reduced mass. Higher orbital $l$ correlations can also contribute using $\partial\delta_0/\partial k \to \Sigma_l(2l+1)\,d\delta_l/dk$. Here only $s-$waves will be considered. Ref[13] gives an early application of the Beth/Uhlenbeck expression to nuclear heavy ion physics.

*Interaction energy.* The volume dependence of the energy can be obtained from $\partial E/\partial V)_T = T\partial P/\partial T)_V - P$. For $P = k_BT(A/V - \hat{b}_2 A^2/V^2)$ with $\hat{b}_2/V = b_2$ the volume dependence of the energy is $E(V) = T(\partial\hat{b}_2/\partial T)k_BT(A^2/V)$ where

$$\hat{b}_2 = -\frac{\lambda_T^3}{2^{7/2}}(y^2 + (1-y)^2) + \sum_S\sum_{i\&j=p,n}y_iy_j\frac{2S+1}{2^2}2^{3/2}\lambda_T^3[\sum_B\exp(\frac{E_B(ij,S)}{k_BT}) + \frac{1}{\pi}\int\frac{\partial\delta_0(ij,S)}{\partial k}\exp(-\hbar^2k^2/mk_BT)dk)] \equiv \hat{b}_{2,sym} + \hat{b}_{2,int} \quad (3)$$

The above volume dependence comes from anti-symmetrization (the term involving $1/2^{7/2}$ or $\hat{b}_{2,sym}$) and from interaction terms (terms with $E_B, \partial\delta_0/\partial k$ or $\hat{b}_{2,int}$) with the latter called the interaction energy. The interaction energy density is

$$\varepsilon_{int} = \frac{E_{int}}{V} = \frac{3}{2}k_BT\frac{A^2}{V^2}\lambda_T^3\sum_S\sum_{i\&j=p,n}y_iy_j\frac{2S+1}{2^2}2^{3/2}\left(-B_{B,C}(ij,S) + \frac{2}{3}T\frac{\partial B_{B,C}(ij,S)}{\partial T}\right) \quad (4)$$

where $B_{B,C}(ij,S) = B_B(ij,S) + B_C(ij,S)$ with

$$B_B(ij,S) = \sum_B\exp(\frac{E_B(ij,S)}{k_BT}), \quad B_C(ij,S) = \frac{1}{\pi}\int\frac{\partial\delta_0(ij,S)}{\partial k}\exp(-\hbar^2k^2/mk_BT)dk \quad (5)$$

A rescaled energy density is defined as $\hat{\varepsilon}_{int} \equiv \varepsilon_{int}/((3k_BT/2)(A/V)^2\lambda_T^3 2^{3/2}/4)$. The experimental determination of the interaction energy in the unitary limit in atomic systems can be found [2,3] and a calculation of it is in ref [7] at non-zero $T$. At $T=0$, a numerical coefficient $\xi$ relating the energy in the unitary limit to the non-interacting Fermi gas energy is of interest. Specifically, $E/N = \xi(3E_F/5)$ where $\xi \approx .3-.4$ [8,11].

*Square well potential with a repulsive hard core: effective range approximation.* The nuclear force has a short range repulsive part besides an attractive longer range part. A simplified interaction with an infinite repulsive core for $0 \le r \le c$ and an attractive square well of depth $V_0$ for $c \le r \le R$ has $\delta_0 = \arctan[(k/\alpha)\tan\alpha(R-c)] - k(R-c) - kc$. The $\alpha^2 = k^2 + \alpha_0^2$ and $\alpha_0 = \sqrt{2\mu V_0/\hbar^2}$. When the square well has no bound state then $\delta_o = 0$ at $k=0$ and reaches a maximum value $\delta_{0,m} = \pi/2 - \sqrt{\pi^2/4 - (2\mu V_0\hat{R}^2/\hbar^2)}$ at $k\hat{R} = k_m\hat{R} = \pi/2 - \delta_{0,m}$, $\hat{R} = R - c$. For a well with a single bound state $\delta_o = \pi$ at $k=0$.

When $k \ll \alpha_0$ the behavior of the phase shift is given by an effective range theory. Specifically, $k \cot \delta_0 = -(1/a_{sl}) + r_0 k^2/2$ with $a_{sl} = R(1 - \tan\alpha_0(R-c)/\alpha_0 R)$ the scattering length and $r_0$ the effective range (see Table1), and $\partial \delta_0 / \partial k$ given by

$$r_0 = R - \frac{1}{\alpha_0^2 a_{sl}} - \frac{1}{3}\frac{R^3}{a_{sl}^2} + c\left(1 - \frac{2R}{a_{sl}} + \frac{R^2}{a_{sl}^2} + \frac{1}{\alpha_0^2 a_{sl}^2}\right)$$

$$\frac{d\delta_0}{dk} = -\frac{a_{sl}}{1 + a_{sl}(a_{sl}-r_0)k^2 + \frac{1}{4}(r_0 a_{sl})^2 k^4}(1 + \frac{r_0 a_{sl}}{2}k^2) \quad (6)$$

Table1. Experimental values of $a_{sl}$ & $r_0$ [19] followed by square well parameters and calculated values. A $c = 0.27$ was used throughout. The np S=1 parameters bind the deuteron at $2.2\,MeV$. The p-p system has Coulomb terms and has $a_{sl} = -7.821$, $r_{0s} = 2.83$.

|  | np  S=1 | np  S=0 | nn  S=0 |
|---|---|---|---|
| Exp: | $a_{sl} = a_t = 5.4$ | $a_{sl} = a_{s,np} = -23.7$ | $a_{sl} = a_{s,nn} = -17.4$ |
| Exp: | $r_0 = r_{0t} = 1.75$ | $r_0 = r_{0s,np} = 2.73$ | $r_0 = r_{0s,nn} = 2.4$ |
| $\{V_0, R\}:$ | {57.14, 1.8} | {23.18, 2.3} | {31.60, 2.0} |
| Cal: | $a_{sl} = a_t = 5.4$ | $a_{sl} = a_{s,np} = -23.70$ | $a_{sl} = a_{s,nn} = -17.4$ |
| Cal: | $r_0 = r_{0t} = 1.73$ | $r_0 = r_{0s,np} = 2.69$ | $r_0 = r_{0s,nn} = 2.40$ |

When the $k^4$ term in Eq.6 is neglected, the integral $B_C$ can be done analytically:

$$B_C \equiv \frac{1}{\pi}\int_0^\infty \frac{d\delta_0}{dk}\exp(-bk^2)dk =$$

$$-\frac{a_{sl}^2(2a_{sl}-3r_0)}{4(a_{sl}^2-a_{sl}r_0)^{3/2}}\exp(\frac{b}{a_{sl}^2-a_{sl}r_0})Erfc(\sqrt{b/(a_{sl}^2-a_{sl}r_0)}) - \frac{a_{sl}^2 r_0}{4(a_{sl}^2-a_{sl}r_0)\sqrt{\pi b}} \quad (7)$$

The Boltzmann exponential factor $\exp(-bk^2)$ suppresses the $k^4$ term in $\partial\delta_0/\partial k$ and makes eq[7] and very good approximation to the complete effective range result for it. The $b = \hbar^2/2\mu k_B T = \lambda_T^2(\mu)/2\pi$ and $Erfc(\sqrt{b/a_{sl}^2}) = 1 - Erf(\sqrt{b/a_{sl}^2})$. Calculations of $B_C$ using a square well $\delta_0$ and an effective range approximation are shown in Fig.1.

Fig.1. $B_C$ versus $b$. Right figure: The upper curve is the approximate analytic expression of Eq. 7 while the lower curve is the square well result. The error is less than a percent. Left figure: unitary limits showing that the two lines nearly coincide.

Fig.1

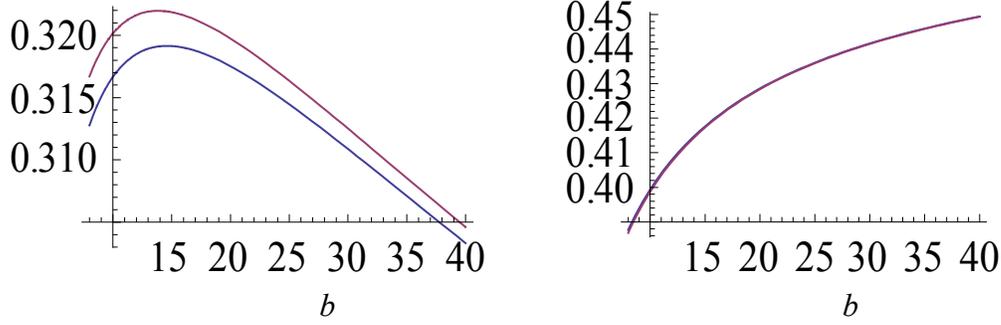

*Unitary or Universal thermodynamic limit.* The limit $a_{sl} \to \infty$ is called the universal thermodynamic or unitary limit. In this limit the scattering length no longer appears in expressions such as the energy, but the energy is also no longer an ideal Fermi gas result. The quantity $\Delta B_C \equiv -B_C - (2/3)b\partial B_C/\partial b$ appears in $\varepsilon_{int}$ and $\hat{\varepsilon}_{int}$. For large $a_{sl}$

$$\Delta B_C \to \frac{1}{6}\frac{r_0}{\sqrt{\pi b}}(1+\frac{r_0}{a_{sl}}+(\frac{r_0}{a_{sl}})^2)+sign(a_{sl})(\frac{1}{2}-\frac{4}{3\sqrt{\pi}}\frac{\sqrt{b}}{a_{sl}}(1+\frac{r_0}{2a_{sl}})+\frac{5}{6}\frac{b}{a_{sl}^2}-\frac{3}{16}(\frac{r_0}{a_{sl}})^2)$$

$$\Delta B_C (a_{sl} \to \infty) \to sign(a_{sl})(\frac{1}{2})+\frac{1}{6}\frac{r_0}{\sqrt{\pi b}} = sign(a_{sl})(\frac{1}{2})+\frac{1}{6}\sqrt{2}\frac{r_0}{\lambda_T} \qquad (8)$$

The factor $r_o/\lambda_T$ appears as a correction to the universal thermodynamic limit of $sign[a_{sl}]/2$. By comparison, for $b/(a_{sl}^2-a_{sl}r_0) \gg 1, B_C \to -a_{sl}/2\sqrt{\pi b}$, $\Delta B_C = -2B_C/3$. The unscaled interaction density $\varepsilon_{int} = a_{sl}(\pi\hbar^2/m)(A/V)^2$ for this $\Delta B_C$ which is realized in atomic systems by tuning away from the Feshbach resonance[7]. The value of $\Delta B_C$ using phase shifts calculated with a square well/hard core potential also closely approximate results from Eq.(8). When $a_{sl} \to \infty$ the effective range $r_0 \to R+c$. A bound state has $-B_b+(2/3)T\partial B_b/\partial T = -Exp[E_b/k_BT](1+(2/3)E_b/k_BT) \to -1$ in the unitary limit of $E_b \to 0$. Thus, in this limit, a bound deuteron has $a_{sl} \to +\infty$ and a contribution $-1+\Delta B_C(a_{sl} \to \infty) = -1+(1/2)+r_0/6(\pi b)^{1/2} = -1/2+r_0/6(\pi b)^{1/2} = \Delta B_C(a_{sl} \to -\infty)$, i.e., the same contribution for an unbound state where $a_{sl} \to -\infty$, assuming that the effective range $r_0$ is the same. The second virial coefficient becomes

$$\frac{\hat{b}_2}{\lambda_T^3} = \{-\frac{1}{2^{7/2}}+\frac{1}{2^{3/2}}-\frac{2^{3/2}}{4}(\frac{1}{4\sqrt{\pi}}\frac{r_{os}}{\sqrt{b}})\}+y(1-y)\{\frac{5}{2^{5/2}}+\frac{2^{3/2}}{4}(\frac{1}{4\sqrt{\pi}}\frac{r_{0s}-3r_{0t}}{\sqrt{b}})\} \qquad (9)$$

The $\hat{b}_2$ includes anti-symmetry exchange correlations and the above formula is for an isospin symmetry case with all $nn, pp, np$ singlet effective ranges ($\equiv r_{0s}$) taken to be the same. The $r_{ot}$ is the $np$ triplet effective range. Figure 2 shows plots of $\hat{b}_2/\lambda_T^3$ versus $y$ in the universal thermodynamic limit and compares it to $\hat{b}_2/\lambda_T^3$ using the effective range and scattering length parameters of Table 1. The EOS is $PV/Ak_BT = 1-(\hat{b}_2/\lambda_T^3)\cdot(A/V)\cdot\lambda_T^3$. The importance of the bound state can be seen in the height in the left figure curves compared to those in the right figure. The small spread in the three unitary limit curves in the left fiqure come from $r_0/\sqrt{b} \sim r_0/\lambda_T$ terms.

Figure 2. $\hat{b}_2/\lambda_T^3$ versus $y$. The left figure is the behavior of $\hat{b}_2/\lambda_T^3$ with $y$ for an isospin symmetric case in the universal thermodynamic limit. A triplet $r_{0t} = 2.03$ and singlet $r_{0s} = 2.67$ was used. The right figure uses experimental $a_{sl}, r_0, E_B$ parameters. The three curves are for $k_BT = 2.59, 3.45, 5.17$ with higher curves corresponding to lower $k_BT$.

Fig.2.

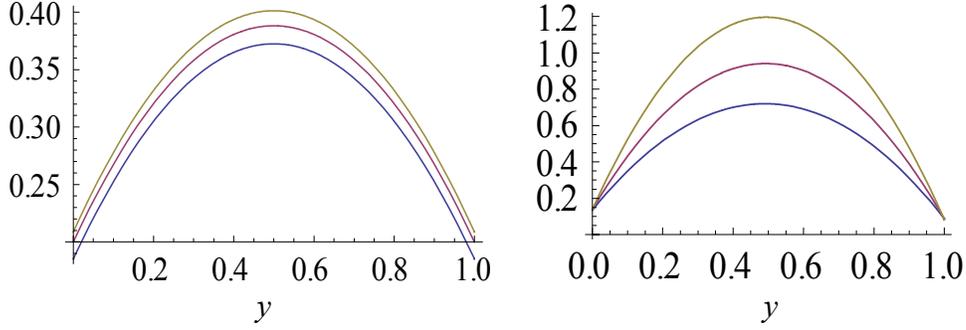

This EOS can be compared with a spinless hard sphere Bose gas (from $l = 0$) [17] and a "spinless" hard sphere Fermi gas (from $l = 1$) [18] having, respectively, $\hat{b}_2/\lambda_T^3 = (1/2^{5/2} + 2c/\lambda_T)$, $\hat{b}_2/\lambda_T^3 = (-1/2^{5/2} + 6\pi(c/\lambda_T)^3)$. A system of protons and neutrons has terms that arise from fermions of the same type - $pp$ & $nn$ - coupled to $S = 0$ for $l = 0$ and fermions that are different - $np$ which can couple to both $S = 0$ & $1$. The rescaled $\hat{\varepsilon}_{int}$ versus $y$ (see Fig.3,4) in the unitary limit is

$$\hat{\varepsilon}_{int} = \{-\frac{1}{2}+(\frac{1}{6\sqrt{\pi}}\frac{r_{os}}{\sqrt{b}})\} + y(1-y)\{-1+(\frac{1}{6\sqrt{\pi}}\frac{3r_{0t}-r_{0s}}{\sqrt{b}})\} \tag{10}$$

The $r_0/\sqrt{b} \sim r_0/\lambda_T$ part of rescaled $\hat{\varepsilon}_{int}$ leads to a $T$ independent $\varepsilon_{int} \sim T\lambda_T^3\hat{\varepsilon}_{int} \sim r_0$ part.

Fig.3 $\hat{\varepsilon}_{int}$ versus $y$ for various $T$. The vertical axis is $\hat{\varepsilon}_{int}$ in $MeV$. The temperatures are $k_BT = 2.59$ ($b = 16.$), $3.45$ ($b = 12$), $5.17$ ($b = 8$) in the left figure. Deeper curves have lower $k_BT$. The right figure has $k_BT = 2.59$ and shows that the isospin asymmetric case is very close to the isospin symmetric case as $y$ varies. The calculated parameters in Table1 for $a_{sl}$ and $r_0$ where used.

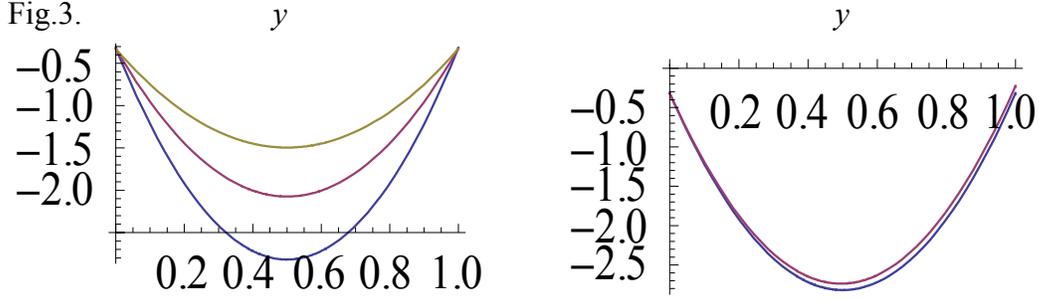

Fig.3.

Fig.4 $\hat{\varepsilon}_{int}$ versus $y$, the role of a deuteron bound state and $T$ dependences. The left figure has $k_B T = 3.45$ or $b = 12$ and the right figure also includes $b = 4$ or $k_B T = 10.15$ to illustrate the role of $T$. In the left figure, the upper curve does not have a deuteron bound state. Instead $a_{sl} = -23.7$, $r_0 = 2.73$ was used for it. The deepest curve has a deuteron bound state as in Fig.3. The middle line is the unitary limit with $E_B = 0$ which is also the same as the unitary limit with an unbound or virtual deuteron. In the right figure, the upper curve is really two curves that almost coincide showing a $T$ independence. The middle two curves are the unitary limits with the lower of the two having lower $T$. The lowest two curves have a bound deuteron with the lower curve having a lower $T$.

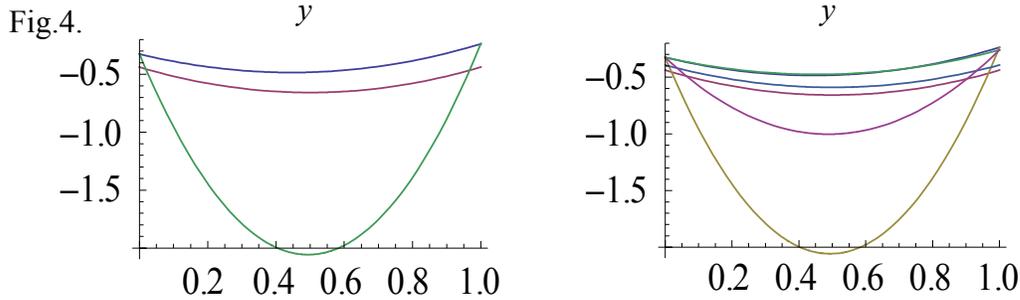

Fig.4.

*Conclusions.* A finite temperature two component model of strongly correlated protons and neutrons each with two spin states and underlying isospin symmetries was discussed. The model is an extension of the one component two spin state fermionic models in both atomic systems and in nuclear physics where pure neutron systems are considered. Features associated with Feshbach resonances and bound states can be studied by tuning on the proton fraction $y$ and varying the temperature $T$. The bound state is the spin $S = 1$, isospin $I = 0$ state of the *np* system, the deuteron. The resonant like virtual structures arise in the $S = 0$, $I = 1$ channels. The *nn* and *np* $S = 0, I = 1$ channels have very large scattering lengths. When the scattering lengths $a_{sl}$ are large compared to interparticle spacings and range of interparticle forces, a regime called the unitary limit is reached and this limit was studied and compared to calculation using experimental values for $a_{sl}$ & $r_0$. A simplied interaction between nucleons was used which is an attractive square well potential with a short range hard core repulsion. Properties of this potential were then related to experimental results for nucleon-nucleon effective ranges $r_0$ and $a_{sl}$ in spin singlet and triplet states. Analytic results were developed in an effective range approximation for various features such as the interaction energy and EOS. A rescaled interaction energy $\hat{\varepsilon}_{int}$ was shown to be relatively flat with variation with $y$ in the unitary limit. A variation with $T$ in $\hat{\varepsilon}_{int}$ comes from a residual dependence on the ratio $r_0 / \lambda_T$ that appears in the results even in the limit of infinite scattering length. The associate $\varepsilon_{int}$ is $T$ independence. The deuteron bound state was shown to produce a large departure in the interaction energy from the unitary limit and giving rise to a pronounced $T$ dependence. Another extreme of large $\lambda_T / a_{sl}$ was also studied. In atomic systems this other limit is realized by tuning away from the Feshbach resonance.

This work supported by Department of Energy under Grant DE-FG02ER-40987.